# Large-scale chirality in an active layer of microtubules and kinesin motor proteins

*Kyongwan Kim[1], Natsuhiko Yoshinaga[1,2], Sanjib Bhattacharyya[1,3], Hikaru Nakazawa[4], Mitsuo Umetsu[4], and Winfried Teizer\*[1,5,6]*

[1]WPI-Advanced Institute for Materials Research (WPI-AIMR), Tohoku University, Sendai 980-8577, Japan

[2]MathAM-OIL, AIST, Sendai 980-8577, Japan

[3]Department of Pharmaceutical Science and Chinese Traditional Medicine, Southwest University, Chongqing 400716, China

[4]Department of Biomolecular Engineering, Graduate School of Engineering, Tohoku University, Sendai 980-8579, Japan

[5] Department of Physics and Astronomy, Texas A&M University, College Station, TX 77843-4242, USA

[6] Materials Science and Engineering, Texas A&M University, College Station, TX77843-3003, USA

# ABSTRACT

During the early developmental process of organisms, the formation of the left-right laterality requires a subtle mechanism, as it is associated with other principal body axes. Any inherent chiral feature in an egg cell can in principal trigger this spontaneous breaking of chiral symmetry. Individual microtubules, major cytoskeletal filaments, are known as chiral objects. However, to date there lacks convincing evidence of a hierarchical connection of the molecular nature of microtubules to large-scale chirality, particularly at the length scale of an entire cell. Here we assemble an in-vitro active layer, consisting of microtubules and kinesin motor proteins, on a glass surface. Upon inclusion of methyl cellulose, the layered system exhibits a long-range active nematic phase, characterized by the global alignment of gliding MTs. This nematic order spans over the entire system size in the millimeter range and, remarkably, allows hidden collective chirality to emerge as counterclockwise global rotation of the active MT layer. The analysis based on our theoretical model suggests that the emerging global nematic order results from the local alignment of MTs, stabilized by methyl cellulose. It also suggests that the global rotation arises from the MTs' intrinsic curvature, leading to preferential handedness. Given its flexibility, this layered in-vitro cytoskeletal system enables the study of membranous protein behavior responsible for important cellular developmental processes.

Before fertilization, a spherical egg undergoes symmetry breaking processes, such as polarity formation and sperm entry, by interacting with the external environment. These pre-developmental processes initiate inhomogeneity in the arrangement of internal materials of the egg, as it enters the developmental process. During this early development, this internal inhomogeneity is responsible for outlining the overall body plan.[1] In the formation of body axes, the left-right laterality is particularly delicate, as it relates to the pre-determined anterior-posterior and dorsal-ventral axes.[2] Intrinsically chiral cellular objects can determine this matter,[2-4] especially when they are associated with a reference membrane.[3] The acto-myosin system has been related to chiral behavior at the whole cell-level.[5-7] Similar to actin, microtubules (MTs) are cytoskeletal chiral proteins,[8-14] responsible for intracellular dynamics.[1] While their involvement in the chiral torsion of the Xenopus egg is still obscure,[5] recent genetic studies suggest an intrinsic role of MTs in the left-right laterality determination,[15] which may be quite universal in-vivo.[16-17] However, no direct evidence has yet been found for large-scale chirality emerging from a system of MTs, driven by kinesin motor proteins. Here we present in-vitro fluorescence microscopy studies of an active layer of MTs and kinesin motor proteins displaying the intriguing emergence of chirality over the entire observed area in the millimeter range. Furthermore, we show that computational models point to spontaneous curvature and local alignment as the reason for the emerging global chirality. This study elucidates the role of MTs in cellular processes accompanied by macroscopic chirality and supports the emergence of the left-right laterality. Compared to acto-myosin-based mechanisms, the MT-kinesin chirality emerges from the directional motility and the molecular handedness only and therefore may be more fundamental, as it does not require special configurations of multiple proteins.[6] Such simplicity facilitates quantitative studies, as well as modeling of the protein-originated chiral development.

**RESULTS AND DISCUSSION**

Quasi-two-dimensional active layers, consisting of MTs and quantum dot-assisted cross-linked kinesin motors (CLKMs), were formed on opposing surfaces of a glass-based flow cell upon sequential introduction of these proteins (details in Materials and Methods). Figure 1a shows the schematic setup, including the definition of normal and clockwise, on both surfaces. The protein density of the active layer was enhanced by repeated protein injections (Fig. 1b). At the lowest density, we find well-isolated, loosely coupled MTs and typical kinesin activity (Fig. 1b top row). Increasing protein density produces a bundled, mesh-like MT network which then evolves with time in a complex manner (Fig. 1b middle and bottom rows). We altered the MT network morphology with methyl cellulose (MC), which allows observation of the kinesin-driven collective behavior of MTs.[18] With increased MC concentration, MTs display wavy bundles aligned along a predominant orientation, as observed previously.[18] This MT alignment persists while the MTs maintain kinesin-driven sliding-like movements, thus generating 'active' aligned bundles (Fig. 2a and Supplementary Movies 1 and 2), qualitatively identical on the top and bottom surfaces. The active MTs are aligned even at an area of $400 \times 400$ μm$^2$ (Fig. 2b). We quantify the degree of alignment by diluting 'visible' MTs, using a mixture of fluorescently labeled and unlabeled MTs (Materials and Methods, Fig. 2c and Supplementary Movies 3 and 4). We then estimate the coherency (degree of alignment) of the MT orientation over a square-shaped region of interest (ROI),[19] using OrientationJ[20] (ImageJ plug-in, https://imagej.nih.gov/ij/, Materials and Methods). Figure 2c shows the coherency becoming less sensitive with increasing area of the ROI and increasing with MC concentration (additional discussion in Materials and Methods), clearly indicating, that the MTs' long-range alignment is stabilized by MC. Surprisingly, this 'active' nematic order extends across the entire cell chamber and is only locally disturbed by defects, such as air bubbles (Figure 2d). The long-range order appears independent of the system size tested (channel width: 1.0 mm ~ 1.8 mm).

The surprising core observation of this work is apparent in the raw data of Figure 2: the global orientation of the active layer rotates with time. Figure 3a shows normalized histograms of local orientations of MTs evolving with time, for four different MC concentrations (Materials and Methods). Initially, the peaks are closely aligned with the flow channel axis, due to the sequential injection of proteins. In the absence of MC, this initial alignment disperses too fast for further quantification (Fig. 3a, far left panel). This dispersion slows down in the presence of MC, allowing the observation of the global rotation with time (Fig. 3a, remaining panels). The variance (Fig. 3d) is decreasing and maintained longer with increasing MC concentration, implying that MC helps the active layer sustain directional order. The global rotation of the active MT layer does not involve any rotation center but is characterized as a self-generated 'shear-like' deformation in a nematic phase. Gliding MTs globally rotate the layer by synchronically rotating locally in the entire device, as long as the MTs are aligned, regardless of their individual polar-orientation (i.e. gliding direction). The most remarkable finding is the universally counterclockwise rotation direction with respect to the surface normal (Fig. 2b). Note that the mean orientation development in Figure 3 also indicates aligned objects globally rotating 'counterclockwise'. Consistent with that, all MT layers formed on a 'top' glass surface show 'clockwise' rotation in the CCD images (Fig. 2a bottom panels and, e.g., Supplementary Movies 2), the 'inverse' of the actual rotation direction (Fig. 1a). Moreover, we observe the chiral rotation of gliding MTs even around the bubble defects. This unidirectional lamellar rotation is direct evidence of large-scale broken symmetry, revealed by the quasi-two-dimensional self-driven many-body dynamics of chiral proteins, entering into a specific structural phase of aligned wavy bundles. The MC dependent rotation of the mean orientation (Fig. 3b) is reproduced in Figure 3c by numerical simulations described in the next section.

In order to clarify the mechanism of the emergence of global chirality, we have developed a theoretical model and performed numerical simulations, outlined in the Materials and Methods. The model consists of motile filaments, each of which has bending rigidity and spontaneous curvature which gives rise to rotation of isolated filaments with a preferential direction. The collective behavior of the filaments appears from the interplay between these properties of single filaments and their interactions. We consider two generic types of interactions, which principally align two filaments during a collision: steric excluded volume interactions and alignment interactions (Supplementary Fig. 1). The alignment interaction literally aligns the polarity of two nearby monomers (Eq.(10) in Materials and Methods). The steric interaction also results in alignment due to Onsager's mechanism[21] (Supplementary Fig. 1) and has been considered as a major source of alignment for active filaments. Hereafter, we shall discuss that this does actually not occur in the present system. While the steric interaction results in inhomogeneous density bands, even in the nematic state (Fig. 4(a)), the alignment interaction shows uniform global polar order (Fig. 4(b)), as seen in our experiments. This suggests that the interactions between microtubules in our experiments are dominated by alignment interactions and not by steric interactions. In fact, we have observed in the experiments that two filaments cross each other over a broad range of incident angles (Supplementary Fig. 2). The inhomogeneous density is understood by motility-induced phase separation which occurs due to the coupling of active self-propulsion and the excluded volume effect.[22, 23] The alignment interaction without the excluded volume interaction does not exhibit phase separation and density inhomogeneity.

We have observed global nematic rotation only for the alignment interactions (Fig. 4(c, d), Supplementary Fig. 10, and Supplementary Movies 5 and 6), when allowing for nonzero spontaneous curvature. The decrease of the mean angular velocity from $3.9 \times 10^{-2}$ $[u/a]$ (see Materials and Methods) to $0.1 \times 10^{-3}$ for $\theta_0 = 10^{-1}$ is significant for the steric interaction, where it almost vanishes in the ordered state. This is in contrast to the alignment interaction, for which the mean angular velocity has a finite value $2.0 \times 10^{-3}$ $[u/a]$ even in the ordered phase, because the rotation of a single filament is incompatible with steric interactions. In order to be aligned by steric interactions, the colliding filaments need to be straight. This inhibits global nematic rotation for the steric interactions. The alignment interaction, on the other hand, occurs at a local scale of monomers, and therefore, survives under nematic order, although suppressed. This is in contrast with vortex formation, that has been discussed in chiral filaments with larger spontaneous curvature[24] and memory effects[25]. Both the nematic order and global angular velocity are dependent on intrinsic properties of active filaments, such as bending rigidity and spontaneous curvature. The global nematic rotation is enhanced at larger bending rigidity (Supplementary Fig. 3(a)). Spontaneous curvature is necessary to obtain global rotation. Nevertheless, too large a spontaneous curvature eliminates the global nematic order (Supplementary Fig. 3(b)).

Does the large scale chirality arise intrinsically from gliding MTs or from an extrinsic factor like the added constituent MC? Analyzing the motion of individual MTs on a CLKM-coated surface (Materials and Methods), we find the mean gliding speed (~ 0.5 μm) of MTs to be independent of the MC concentration (Fig. 4e), while the mean angular velocity systematically decreases from about 0.3 degree/sec to below 0.05 degree/sec with increasing MC concentration (Fig. 4f). Importantly, the mean rotational direction is counterclockwise, consistent with our prior finding. This symmetry-broken angular velocity increases with decreasing MC concentration, is even present in the absence of MC, and is therefore caused by an intrinsic property of gliding MTs. This biased gliding of MTs then results in the chiral rotation upon emergence of global alignment, implying the necessity of intrinsic curvature for global rotation. Varying the tubulin polymerization condition alters the number of protofilaments, which in turn changes handedness of the MT 'supertwist'.[8] If the origin of the biased angular velocity was related to the superstructure's handedness, then the large scale rotation would switch from counterclockwise to clockwise, upon a switch of the superstructures' handedness. However, none of the tubulin polymerization conditions we used exhibited such switching. Moreover, an analysis of the gliding motion of individual MTs polymerized in these atypical conditions also indicates preferentially counterclockwise mean angular velocities (Materials and Methods). These results point to the biased rotation arising from structural characteristics of MTs at an even more fundamental level, i.e. the MTs' left-handed lattice. The global tilt of protofilaments with respect to the MT axis preserves this left-handed nature of MTs, even if a lattice mismatch occurs along a 'seam' due to any change of the number of protofilaments.[26] Thus, the basal handedness of MTs can produce an asymmetric twisting rigidity.[27] Additionally, we evaluated two correlations as functions of lag time: the angular velocity and the local gliding orientation of MTs (Materials and Methods). While the angular velocity correlation collapses within our time resolution (5 sec), the orientation correlation lasts for considerable times (~50 – 100 sec), that increase with MC concentration, indicating that MC induces suppression of the orientation fluctuations of gliding MTs.

Throughout all multicellular species, cytoskeletal filaments, paired functionally with motor proteins, are categorized into only two kinds, MTs and actin-filaments. Given this universality, collective movement of these motility constituents may be the most fundamental aspect of the cell to transfer inherited cellular-level 'behavioral' information, that drives the robust growth of living organisms. Cytoskeletal proteins may have built their own structural signaling mechanism governed by intrinsic geometrical codes of these proteins[2-4] or by the fundamental concept of percolation.[28] In light of this, it is enticing that the morphological progression of our in-vitro MT network is reminiscent of that of cortical MTs during the early development of the fertilized Xenopus egg, including aligned wavy bundles.[29, 30] A relation of the cortical rotation process to the emergence of the orderly cortical MT arrangement, known to be critical for body axis formation, has been suggested:[31-33] Aligned MT layers are associated with molecular motors, such as kinesin and dynein, either anchored on the cortex wall or coupled to MTs within the MT layer. These motor proteins interact with MTs, producing a force to move the MT layer, which in turn generates a global rotation of the cortex of the egg relative to its core. Interestingly, this yet speculative scenario resembles our in-vitro protein configuration in quite some detail. A similar MT arrangement, accompanied by a cortical rotation-like process, was recently identified in Zebrafish eggs as well,[34] suggesting that it is a dynamic phase, common in early development. Moreover, cellular-level rotation of aligned membranous MTs seems to be prevalent across kingdoms of life.[35-37] The clear identification of the intriguing MT dynamics, including the MTs' wavy arrangement, provides promising insights into the yet unknown mechanisms of cellular chirality.

**CONCLUSIONS**

Various *in-vitro* cellular mimetic studies have been performed upon reconstitution of the cytoskeletal protein motility systems. The *in-vitro* approach has allowed the study of active many-body systems, revealing the emergence of various dynamic patterns,[38-41] directly relevant to cellular processes.[42-46] Emphasizing potential connections to cellular systems, we have found new features emerging from collective protein dynamics in a system of MTs and kinesin motor proteins. Increasing system complexity gives rise to a unique dynamic structural phase, active MT bundles aligned in a wavy pattern, reminiscent of the morphology of what has transiently been observed in Xenopus eggs. This directional order is extremely long range, covering our entire experimental system size (millimeters), and is thus comparable to the egg size. Surprisingly, this emerging macroscopic order unveils a hidden broken symmetry, a 'counterclockwise' unitary rotation of the large scale active nematic layer. Analysis of the gliding dynamics of isolated MTs reveals an intrinsically biased counterclockwise angular velocity. Our theoretical model suggests that bending rigidity, spontaneous curvature and interactions between filaments control the global collective behavior. In-vivo, these parameters are controlled by MC but also by other tubulin-related proteins and drug treatments, leading to different local molecular structures. Therefore our novel finding provides a generic recipe for controlling large-scale chirality in cells and explains the hierarchical emergence of the global order of a scaled up MT system from the molecular-level dynamics, including the robust handedness, protected by the collective behavior. In addition, local alignment interactions result in global nematic rotation, which is not seen for steric interactions, raising questions about several inconsistent claims in the literature.

**MATERIALS AND METHODS**

*Proteins*

Commercial tubulin proteins were polymerized into MTs in a standard polymerization buffer (PEM buffer (80 mM PIPES (P6757, Sigma-Aldrich), 1 mM EGTA (E0396, Sigma-Aldrich), 1 mM MgCl$_2$ (M8266, Sigma-Aldrich), pH 6.9 (NaOH controlled)) with 1 mM of GTP and glycerol at 6 % (v/v))[47] at 37 °C for 1 hr at a ratio of 7:3 of unlabeled tubulin (T240, Cytoskeleton) to rhodamine labeled tubulin (TL590M, Cytoskeleton).[48] Kinesin motor proteins were expressed in *Escherichia coli* following a standard protocol.[49] In this study, we used the expression plasmid of a *Neurospora crassa* truncated kinesin heavy chain (400 residues),[50] extended with the additional functional domain for biotin attachment, followed by the hexa-histidin-tag. The overall procedure for the kinesin expression is described in depth elsewhere.[48]

*Glass-based flow cell chamber and fluorescence microscopy*

Each flow cell chamber was built using a glass slide (size: 76 × 26 mm$^2$, thickness: 0.8 ~ 1.0 mm, S2111, Matsunami), a glass coverslip (size: 18 × 18 mm$^2$, thickness No.1: 0.12 ~ 0.17 mm, Muto pure chemical), and two pieces of double-sided tape as spacers (thickness: 30 μm, TL410S-06, Lintec), in a typical flow channel geometry as shown in Fig. 1. Fluorescence microscopy used an inverted optical microscope (IX71, Olympus), equipped with a CCD camera (ImagEM, Hamamatsu) and a filter-set for rhodamine. Two different adaptors (microscope-to-camera: 1x and 4x magnification) and two different lenses (20x and 40x magnification) were used with three different combinations in order to vary the field of view (51.4 × 51.4 μm$^2$, 204.8 × 204.8 μm$^2$ and 409.6 × 409.6 μm$^2$). Movies were recorded at a rate of 2 frames/sec.

*Assay protocol for the active layered MT-kinesin network*

MTs obtained by the tubulin polymerization process were 40× diluted in PEM buffer including 15 μM taxol (T7402, Sigma-Aldrich)), resulting in the final tubulin concentration of 1.13 μM. The quantum dot (QD)-assisted CLKMs were prepared by mixing QDs (we use either QD525 (Q10141MP, Invitrogen) or QD655 (Q10123MP, Invitrogen), stock solution concentration: 1 μM) with kinesin proteins (stock solution concentration: 50 μM) in PEM buffer at the volume ratio of 1:1:8 (kinesin : QD : PEM). The assay protocol is as follow: **1)** a flow cell was filled with casein solution (037-20815, Wako, 2 mg/ml in PEM, used after syringe filtering (0.2 μm, Puradisc, Whatman)) and incubated for 10 min at room temperature. **2)** Three-fold further diluted CLKM solution was introduced into the flow cell by fluid exchange using a pipette for solution injection and a piece of filter paper for soaking, followed by 10 min incubation at room temperature. **3)** The cell was flushed with PEM buffer. **4)** MT solution including ~ 5 mM ATP (adenosine tri-phosphate, A9187, Sigma-Aldrich) was introduced into the cell, followed by 10 min incubation at room temperature. **5)** Finally, motility solution (20 mM glucose (076-05705, Wako), 10 mM ATP, 20 μg/mL glucose oxidase (G7141, Sigma-Aldrich), 8 μg/mL catalase (C40, Sigma-Aldrich), 0.01 %(v/v) 2-mercaptoethanol (190242, MP-Bio Japan), 10 μM taxol and CLKM (30x final dilution of the starting solution described above) in PEM buffer) was injected into the cell, followed by the fluorescence microscope observation after VALAP (Vaseline-Lanolin-Paraffin) sealing. In order to increase the MT and CLKM densities in the active layer, two additional steps, CLKM solution injection (20x final dilution of the starting solution in 15 μM taxol-included PEM buffer) and MT solution (with ATP) injection, were sequentially performed before the motility solution injection. Incubation time for these additional protein injections was 3 min and 5 min, respectively. For further enhancement of the protein densities, we repeated these two additional steps once more before the motility solution injection. Three different repetitions of MT solution injection, single, double and triple, were performed in this study (see Fig. 1b).

*MC-induced aligned active layer*

Assay protocol for the aligned active layer is essentially the same as described above for the active network formation, except for the MT density 2-fold increased (final tubulin concentration: 2.26 μM) and the inclusion of MC in the motility solution. The two additional steps for protein density enhancement described above were repeated twice (*i.e.* triple injection of MT solution). The final concentration of MC (M0512, viscosity: 4000cp, Sigma-Aldrich) in the motility solution to achieve the dynamic alignment of the MT-CLKM layer in Figure 2 (except panel c) was 0.34 % wt. In general, we monitored the bottom glass surface for 10 min, and then switched the focal plane to the top glass surface for another 10 min monitoring, and repeated this sequential (bottom surface-top surface) monitoring process for a total time of 40 ~ 60 min for each assay.

*Dilution of visible MTs*

To decrease 'visible' MTs, keeping the overall MT density at the similar level, we polymerized two tubulin stocks. One was the same labeled tubulin (stock concentration 5 mg/ml) as mentioned above (including 30% of rhodamine labeled tubulin) and the other one was fully unlabeled tubulin (stock concentration 5 mg/ml). These two different MT solutions (*i.e.* after polymerization) were then mixed at a ratio of 1:25 ~ 1:130, thus controlling the degree of separation between visible (labeled) MTs.

*Coherency measurement*

Four different MC concentrations including the case of no MC were analyzed. The final concentration of MC in the motility solutions were, 0, 0.11, 0.34 and 0.56 % wt, referred to as MC0, MC10, MC30 and MC50, respectively. The coherency of MT orientations was measured using OrientationJ. A region of interest (ROI) was defined by a square and the coherency was measured over the ROI, before the region was changed for the next measurement. In this coherency measurement mode (OrientationJ Measure option), OrientationJ Measure uses a weight factor uniform over the whole area of ROI. Thus it evaluates mean coherency across the ROI.[20] This area scan was performed covering the entire image from the left-top corner to the bottom-right corner without overlap. Six different areas of the ROI were selected: 16 × 16, 32 × 32, 64 × 64, 128 × 128, 256 × 256 and 512 × 512 pixel. The length-pixel conversion factor was 0.8 μm/pixel (E.g. 512 × 512 pixel corresponds to the area of 409.6 × 409.6 μm$^2$). We calculated the mean coherency for each ROI by taking the average over the multiple data for each of the six different areas (except for the case of 512 × 512, which has only one data point), and plotted it as a function of the area. Seven different time frames were chosen for this analysis, covering the whole period of time from zero to 10 min with an interval of 100 sec. Supplementary Fig. 4(c, d) shows the measured coherency for the two different MC concentrations, which vary with the area of ROI and with time.

*Analysis of orientation evolution*

To estimate temporal mean orientations of the active layers, we first selected seven different time frames, as in the previous case (from zero to 10 min with an interval of 100 sec), and constructed histograms of local orientations for each MT fluorescence image. This was done by using OrientationJ (Parameters: Gaussian window: σ = 1 pixel, Cubic spline gradient, Minimum coherence: 70 %, Minimum energy: 0 %). Resulting histograms are shown respectively in Supplementary Fig. 4e and f for two different MC concentrations. Finally, we defined the mean orientation for each time frame as the center of the Gaussian distribution, fitting the normalized histogram (the histogram data divided by the maximum count). This mean-orientation analysis was done only for the three cases of finite MC concentrations. The variance of orientation distribution, $\langle (\delta\theta)^2 \rangle$, was calculated by the formula,

$$\langle (\delta\theta)^2 \rangle = \frac{\sum_i \{(\theta_i - \theta_{mean})^2 P(\theta_i)\}}{\sum_i P(\theta_i)},$$

where $\theta_{mean}$ is the mean orientation obtained from the Guassian curve fitting to the orientation histogram and $P(\theta_i)$ is the count corresponding to the angle $\theta_i$.

*Analysis of MT dynamics at the individual-level*

To analyze the dynamic properties of individual MTs gliding on a CLKM-coated surface, MT gliding assays (the case of single introduction of MT solution: see the assay protocol described above) were performed with diluted MT concentration (tubulin concentration: 9 ~ 46 nM). Positions of MTs, evolving with time, were extracted by applying an automatized particle tracking program (TOAST, imageJ plug-in)[51] to fluorescence microscope time-lapse movie data (converted to binary formats: see Supplementary Movies 7 for a typical MT gliding assay ). All the analyzed samples were observed at a 409.6 × 409.6 μm² field of view. To reduce counting inter-filament collision events, we removed all the data points of inter-filament distances smaller than 20 μm. The frame size of each original movie (10 min, 1201 frames) were 10 fold reduced (121 frames) for the TOAST analysis. The time interval ($\delta t$) between the reduced frames was therefore 5 sec. A single set of TOAST parameters was applied to all the samples analyzed. First, the translational speed, $\frac{|\mathbf{r}(t+\delta t)-\mathbf{r}(t)|}{\delta t}$, and the angular velocity, $\frac{\sin^{-1}(\mathbf{u_i} \times \mathbf{u_f})}{\delta t}$, were calculated for each traced MT. Here, $\mathbf{r}(t)$ is the position $(x(t), y(t))$ of the traced MT at time $t$, $\mathbf{u_i}$ and $\mathbf{u_f}$ are initial and final moving directions at time $t$, denoted as $\frac{\mathbf{r}(t+\delta t)-\mathbf{r}(t)}{|\mathbf{r}(t+\delta t)-\mathbf{r}(t)|}$ and $\frac{\mathbf{r}(t+2\delta t)-\mathbf{r}(t+\delta t)}{|\mathbf{r}(t+2\delta t)-\mathbf{r}(t+\delta t)|}$, respectively. The mean translational speed was then obtained for each observation by directly taking an average of the translational speeds collected from all the traced MTs. Meanwhile, the mean angular velocity was evaluated as the center of the Lorenztian curve fitting to the angular velocity histogram. Supplementary Fig. 5 shows an example of the histogram and the Lorenztian curve. In this analysis, a negative value of the angular velocity means counterclockwise rotation as the origin of the image frame was set at the top-left corner, and *x* and *y* increases rightward and downward, respectively. In the first trial of this analysis, one assay was performed for each MC concentration, and three different regions for each surface (in the sequence of bottom-top-bottom-top-bottom-top) were observed within each assay. In this trial, we found MT gliding speeds gradually decreasing with time (Supplementary Fig. 6a). This may indicate the interaction between MTs and CLKMs becomes more nonspecific losing their binding with time. In the second trial, we thus increased the numbers of assays to three while observing two different regions for each surface (in the sequence of bottom-top-bottom-top). We performed this second trial only for two cases, MC0 and MC30. Supplementary Fig. 6(b,c) show the results from the second trial.

*Orientation and angular velocity correlation*

Correlation functions were computed using the position data produced by TOAST. The orientation (moving direction) correlation function was defined as $\frac{\langle \mathbf{u}(\Delta t) \cdot \mathbf{u}(0) \rangle}{\langle \mathbf{u}(0) \cdot \mathbf{u}(0) \rangle}$. The angular velocity correlation function was defined as $\frac{\langle \omega(\Delta t) \omega(0) \rangle}{\langle \omega(0) \omega(0) \rangle}$. Here $\Delta t$ is the lagging time and $\langle A \rangle$ is $\sum_{i=1}^{N(\Delta t)} A_i$. Note that $i$ is the MT index and $N$, the total number of counted MTs, is a function of $\Delta t$ because each MT can have a different traced time. First, these correlation functions were computed for each observation (10 min). Then, the final correlation function for each specific experimental condition, such as a MC concentration, was obtained by taking the average of the correlation functions over all independent observations performed at the same specific condition. Supplementary Fig. 7(a,b) shows the correlation functions for four different MC concentrations, the first trial which was analyzed for Fig. 4 and Supplementary Fig. 6a. Supplementary Fig. 7(c,d) shows the correlation functions and exponentially decaying curves fitting the OCFs (orientation correlation functions) with a correlation time, $\tau$, for the case of the second trial, which was analyzed for Supplementary Fig. 6(b,c).

*Dynamics of MTs polymerized under exotic conditions*

**Collective dynamics:** We performed assays following the above mentioned protocol to see the active layer formation and the collective dynamics, using MTs polymerized in various atypical buffer conditions. The primary sets of the performed polymerizaton buffer conditions (with parameter ranges) are categorized as follows: **1)** Standard polymerization buffer adjusted with dimethyl sulfoxide (DMSO: 5% (0.6M) ~ 12.5%), incubation time: 1 ~ 3 hr, tubulin concentration: 2.5 mg/ml. MTs grew poorly as the DMSO concentration was increased and the sample condition was not clean for observation (dot-like aggregates dominant) with longer incubation times ( > 4 hr). **2)** Phosphate buffer (10 mM) including taxol (2 ~ 10 μM), incubation time: 1 ~ 3 hr, tubulin concentration: 0.5 ~ 5 mg/ml. The sample condition was not clean for observation (low MT density and dot-like aggregates dominant) with phosphate butter. **3)** PEM buffer including taxol (10 ~ 80 μM), incubation time 3 ~ 24 hr, tubulin concentration 0.5 ~ 5 mg/ml. **4)** Guanosine-5'-[(α,β)-methyleno] triphosphate (GMPCPP, 1 mM) added PEM buffer or PEM with 6% DMSO, incubation time: 3 hr, tubulin concentration: 0.5 mg/ml (PEM) and 1.0 mg/ml (DMSO-PEM). Our typical polymerization buffer is known to produce predominantly 14-protofilament MTs, which have a supertwist handedness opposite to that of MTs with 12- or 15-protofilaments.[8] Condition 1) was chosen with an expectation of switching the favored number of protofilaments to 15.[11] 2) and 3) were chosen with an expectation of switching the favored number of protofilaments to 12.[8,10] 4) was chosen with an expectation of structural modification of MT.[52] Interestingly, in the case of GMPCPP induced MTs, trajectories of gliding MTs were very straight compare to those from other cases. The GMPCPP-MTs display very slow rotation of the MT layer, yet the rotation direction is still counterclockwise. This slow rotation may reflect the enhanced MT rigidity with GMPCPP buffer.[53] Alternatively, the GMPCPP buffer may decrease the intrinsic curvature of MTs, as pointed out in the modeling analysis (see Supplementary Fig. 3b). **Individual MT dynamics:** Gliding dynamics of isolated MTs was analyzed for several selected cases using TOAST following the same procedure described before. MT gliding assays were performed with highly diluted MT solution (see above Assay protocol and Analysis of MT dynamics at the individual-level). Resulting mean angular velocities are described in Supplementary Fig. 8 for each case with detailed information about the polymerization conditions. Three different cases were chosen for the relaxation time analysis. Like before, we first computed correlation functions (Supplementary Fig. 9(a,b)) and then obtained relaxation times by exponential curve fitting (Supplementary Fig. 9(c-e)). Mean MT gliding speeds are also plotted in Supplementary Fig. 9f indicating no noticeable dependence on the polymerization condition.

*Modeling*

We consider a simplified model of microtubule dynamics. The two-dimensional model consists of $N$ filaments of identical length $2aM$, where each filament consists of $M$ circular monomers with radius $a$. The density of the filaments, $\rho$, is defined by $\rho = \pi a^2 NM/L^2$, where the box size of the system is denoted by $L$ under periodic boundary conditions. The filament has bending rigidity $\kappa$ and a spontaneous curvature $\theta_0 \in [-\pi, \pi]$, so that it forms a curved conformation. The signed spontaneous curvature arises from confinement of a polymer with spontaneous curvature and helical twist in three dimensions onto a two-dimensional surface[54,55] and sets the radius of curvature of the conformation and its sign determines the preferential direction, left ($\theta_0 < 0$) or right ($\theta_0 > 0$). Motility, driven by kinesin motors, is modelled by spontaneous velocity along the local orientation of a filament. Under fluctuations, an isolated filament changes its orientation and, as a result, changes its moving direction. When the bending rigidity is high, the filament prefers a straight shape, so that it moves almost along a straight path for a short time. Therefore, the orientational correlation time is affected by the bending rigidity. Each filament is modelled by $M$ beads connected by a spring (the beads-spring model) and the position, $\mathbf{x}_i(t)$, and orientation, $\mathbf{p}_i(t) = (\cos\phi_i(t), \sin\phi_i(t))$, of the $i$th particle ($i$ is a integer in $[1, NM]$) are updated by

$$\frac{d\mathbf{x}_i}{dt} = \mathbf{v}_i \tag{1}$$

$$\frac{d\phi_i}{dt} = \omega_i - k_r\left(\phi_i - \phi_i^{(0)}\right) \tag{2}$$

$$\frac{d\mathbf{v}_i}{dt} = -\gamma\left(\mathbf{v}_i - u_0\mathbf{p}_i\right) + \sum_{j\neq i}\mathbf{f}_{ij} + \boldsymbol{\xi}_i \tag{3}$$

$$\gamma_r \omega_i = \sum_{j\neq i} g_{ij}. \tag{4}$$

The velocity and angular velocity of a particle are denoted by $\mathbf{v}_i$ and $\omega_i$, respectively. The position and orientation are confined in two dimensions, as they are in the experiment. Here, $\gamma$ and $\gamma_r$ are the translational friction constant between the filament and the substrate and the rotational friction constant, respectively. Gaussian white noise $\boldsymbol{\xi}_i$ with amplitude $\sigma$ is added in (3). The polarity of a monomer is aligned with the mean orientation of local segments

$$\mathbf{p}_i^{(0)} = \left(\cos\phi_i^{(0)}, \sin\phi_i^{(0)}\right)$$

$$\phi_i^{(0)} = \mathrm{Arg}\left[\frac{1}{2}\left(\frac{\mathbf{r}_{i+1}-\mathbf{r}_i}{|\mathbf{r}_{i+1}-\mathbf{r}_i|} + \frac{\mathbf{r}_i-\mathbf{r}_{i-1}}{|\mathbf{r}_i-\mathbf{r}_{i-1}|}\right)\right], \tag{5}$$

with a restoring spring constant $k_r$. Here Arg denotes the argument of a vector inside the parentheses. The force acting on each monomer is expressed by the potential $U$ as

$$\mathbf{f}_{ij} = -\frac{\partial U}{\partial \mathbf{r}_{ij}}. \tag{6}$$

In the beads-spring model, two monomers are connected by a spring with the potential

$$U_{\mathrm{spring}} = \frac{k}{2}\sum_{(i,j)\in\mathrm{neighbor}} (r_{ij} - 2a)^2, \tag{7}$$

where $k$ is the spring constant. In this work, we set $k = 100 k_B T$, so that the relaxation time of the vibration is fast. It is known that a microtubule filament has bending rigidity. The persistence length $l_p = \frac{\kappa}{k_B T}$, which characterizes the rigidity, is about 5000 μm -.[56] The bending rigidity is expressed by the following potential energy

$$U_{\mathrm{bend}} = \frac{\kappa}{2}\sum_i (\theta_i - \theta_0)^2, \tag{8}$$

where $\theta_i$ is the angle between two neighboring segments $\mathbf{r}_{i,i+1}$ and $\mathbf{r}_{i-1,i}$. The filaments are also allowed to have spontaneous curvature, if they prefer a curved conformation. This is characterized by the angle $\theta_0$. The radius of the curvature is $2a/\theta_0$. When $\theta_0 = 0$, the filaments prefer to be straight. When the filament is straight and isolated, it moves at the constant speed $u_0$.

From the experimental results $u_0 \approx 0.5$ m/sec and it is not sensitive to the MC concentrations (Figure 4a) and the tubulin polymerization conditions (Supplementary Fig. 9f).

The interactions between monomers are modelled by steric repulsion (Eq.(9)) and alignment (Eq.(10)). The translational force arises from the potential. The steric interaction is expressed by a soft potential within the radius of a monomer

$$U_{\text{steric}} = \frac{\beta}{2} \sum_{i \neq j} (r_{ij} - 2a)^2 \tag{9}$$

if the distance between the $i$th and $j$th particle, $r_{ij}$, satisfies $r_{ij} \leq 2a$, and $U = 0$ otherwise. We assume that all monomers have the same radius $a$. This steric interaction is soft and thus monomers may overlap. Nevertheless, when the strength of this interaction, $\beta$, is strong, overlap is prohibited. This choice of a soft repulsive interaction is justified from experimental observation, since a filament sometimes crosses another filament. The rotational interaction is given by

$$g_{ij} = \frac{\alpha}{n} \sum_{(i,j) \in r_{ij} \leq 2a} \sin 2(\phi_j - \phi_i), \tag{10}$$

where $n$ is the number of particles within a distance $2a$ from the center of mass of the $i$th particle. This interaction ensures nematic alignment of the orientation of monomers. The strength of the alignment is characterized by $\alpha$. Initial conditions of the simulations are fitted with experiments and are chosen, so that filaments are placed at random positions and are aligned along the $x$-axis with random orientation to the $+x$ and $-x$ directions. We have confirmed that the results are unchanged when we use the initial conditions of random orientation in two dimensions. The correlation function of the angular velocity is given by

$$\langle \omega_i(t)\omega_i(0) \rangle = \langle \omega_i^2(0) \rangle e^{-t/\tau_r}, \qquad (11)$$

where $\tau_r$ is the relaxation time. Supplementary Fig. 7(a,c) (see also Supplementary Fig. 9a) shows that this time scale is fast ($\tau_r < 5$ sec, the resolution limit of our analysis), and therefore, the inertia of the rotation is negligible. We also neglect the inertia term in translational motion in (3). In Fig. 3c, we normalized the length scale by the monomer size $a$ and, using the speed of motility $u$, we normalized the time scale by $a/u$. Using the experimental results of length of a single filament and translational speed, $a \approx 0.16$ μm and $a/u \approx 0.32$ sec.

Without interactions between filaments, $\alpha = 0$ and $\beta = 0$, they move independently, and therefore the system is in a disordered state. For $\alpha/k_B T \gg 1$ or $\beta/k_B T \gg 1$, the system develops nematic order, where filaments are oriented in the same direction. This transition occurs at $\alpha_c \approx 1$ or $\beta_c \approx 2$ (Supplementary Fig. 10). This is quantified in the nematic order parameter

$$S = \frac{1}{NM}\left|\sum_j e^{2i\phi_j}\right|. \qquad (12)$$

When $S < 1/\sqrt{N}$, the system is in the disordered state, whereas the system is in the nematic state when $S \approx 1$. The nematic order induced by the alignment interaction is the same as the transition in the Vicsek model and its variant using point particles.[57-59] On the other hand, when $\alpha = 0$, there is no explicit interaction of alignment, but the excluded volume interaction gives rise to alignment.[60] This is demonstrated by a disorder-nematic phase transition of passive nematic liquid crystals under repulsive interaction between rod-like liquid crystalline molecules.[61] We may also define a mean angle of the system from the argument of a complex-value representation of mean orientation as follows:

$$\langle\phi\rangle = \text{Arg}\left[\frac{1}{NM}\sum_j e^{2i\phi_j}\right]. \tag{13}$$

We measured the mean angular velocity by linear fitting of $\langle\phi\rangle$ as a function of time.

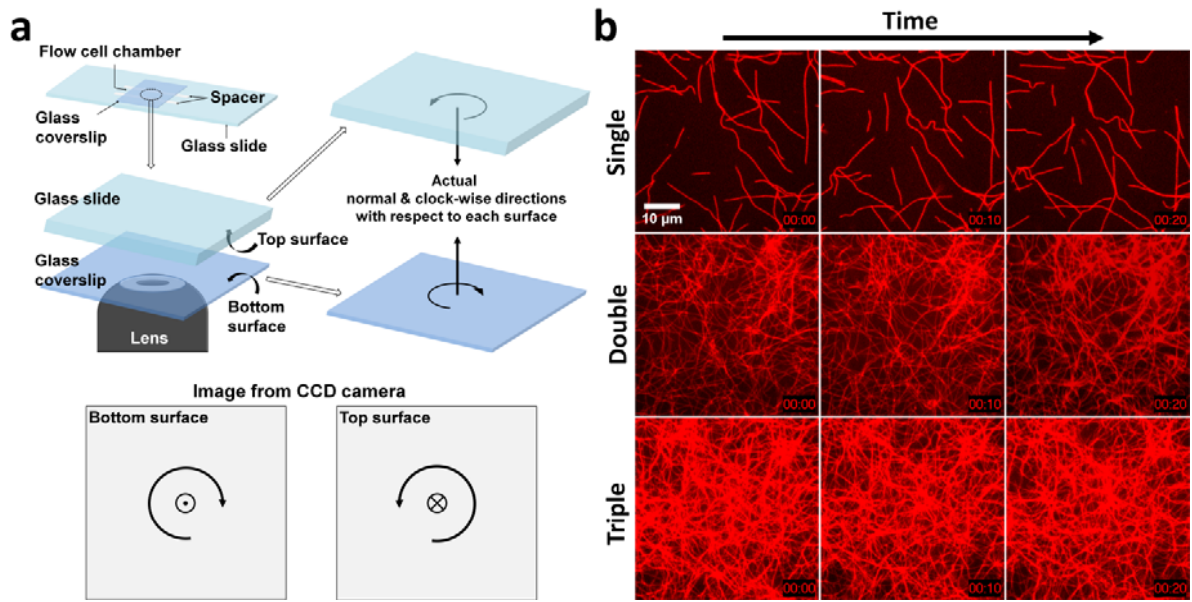

**Figure 1**. In-vitro active layer of MT-CLKMs: **(a)** A schematic diagram of the glass flow cell (top left). The flow cell is mounted upside down in the microscope, as shown in the cartoon right below the flow cell schematic. Two directions, surface normal and clockwise, are defined in the top right diagrams. Note that CCD camera pictures and movies show objects as they are seen from the top, though the lens is located below the objects. Thus the actual 'clockwise' direction with respect to the surface normal defined on the top surface appears 'counterclockwise' in the CCD image. On the contrary, the CCD image for the bottom surface indicates the actual 'clockwise' direction with respect to its surface normal as 'clockwise', without such an inversion. The bottom diagrams depict how the two directions are defined in the images captured by the CCD camera. **(b)** Fluorescence images (after auto bright & contrast adjustment in ImageJ) of MTs evolving with time (time interval: 10 sec, the bottom right time stamp is in min:sec). The scale bar measures 10 μm. Each row represents MT network morphology achieved with different protein density, which was controlled by increasing the number of repetitions of protein injection, from single to triple, respectively (see Materials and Methods). All the images were captured at the bottom glass surfaces.

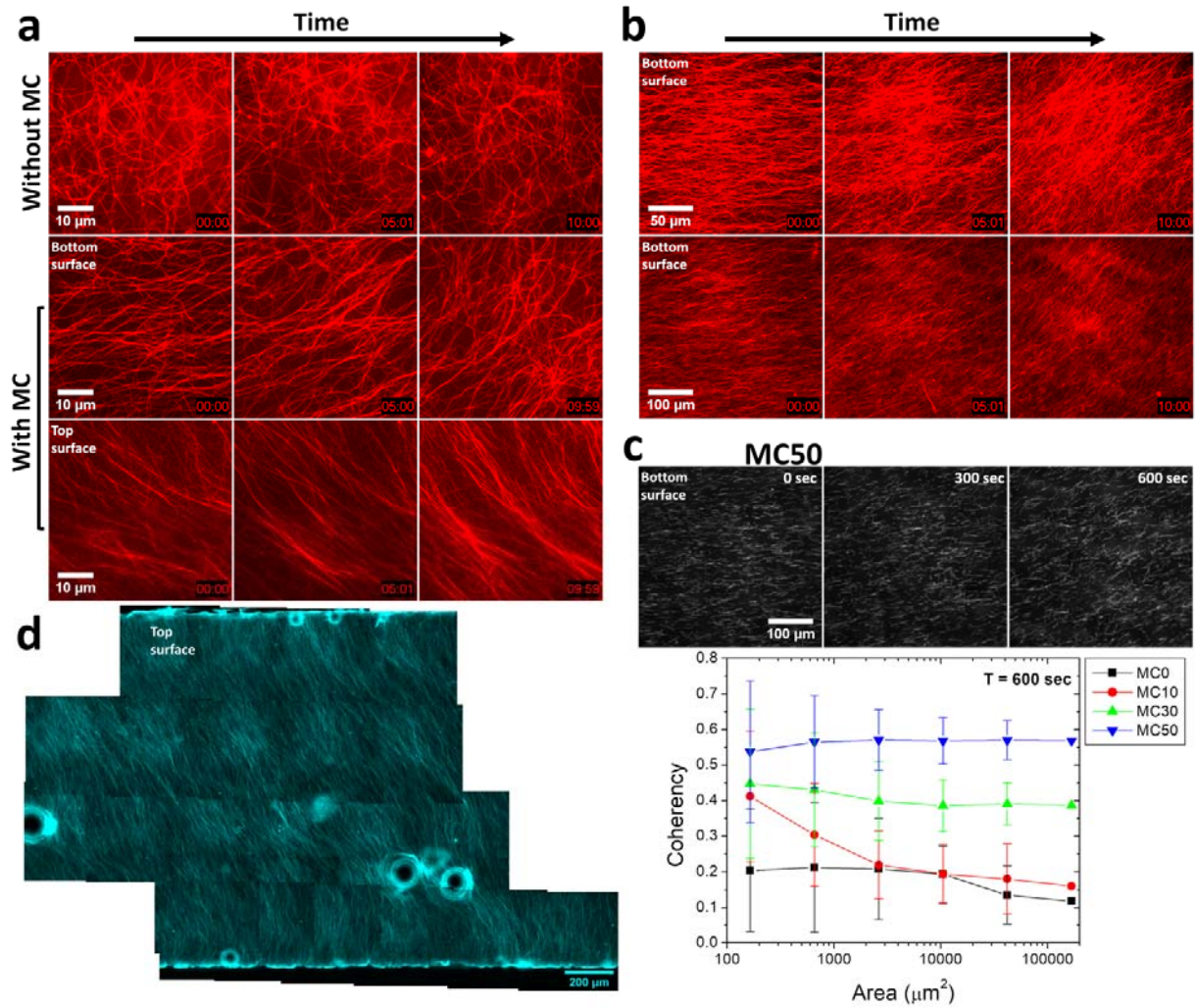

**Figure 2.** Methyl cellulose (MC)-induced long-range directional order of the active layer: **(a)** Fluorescence images of MT networks evolving with time (time interval ~ 5 min, the bottom-right time stamp is in min:sec). The scale bars all measure 10 μm. Top row: MT network formed in the absence of MC. Middle and bottom rows: MT network formed on the top (middle row) and bottom (bottom row) glass surfaces in the presence of MC, displaying MT bundles aligned along a specific orientation that evolves with time. **(b)** Fluorescence images of MT networks evolving with time (time interval ~ 5 min, the bottom-right time stamp is numbers in min:sec): wavy patterns of the aligned active bundles of MTs observed at larger fields of view. The scale bars measure 50 μm (top row) and 100 μm (bottom row), respectively. Each set of the images is from an independent assay. **(c)** Fluorescence images of diluted visible MTs in an aligned active layer evolving with time (scale bar: 100 μm). The bottom graph shows system size-dependent coherency measured after 10 min passed from the beginning of observations for four different MC concentrations. The fluorescence images are for the case of the highest MC concentration (see Materials and Methods). **(d)** A composite image made of several fluorescence sub-images, captured on different areas of an active layer formed on a top glass surface. Circular objects are air bubble defects and line boundaries at the top and bottom are the two sides of the flow cell chamber defined by spacers (see Fig. 1a and Materials and Methods). The scale bar measures 200 μm. Colors were arbitrarily chosen.

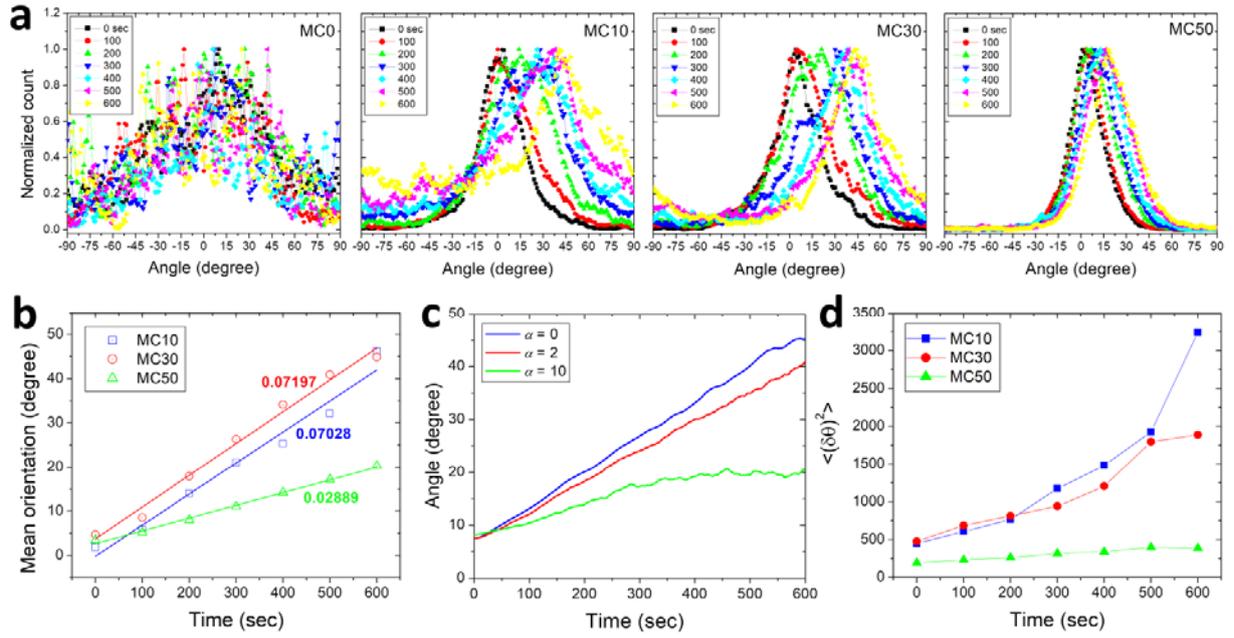

**Figure 3.** Counterclockwise global tilting of the mean orientation: **(a)** Time evolution of normalized orientation histograms (histogram data divided by the maximum count for each plot) measured on MTs distributed in active layers formed on bottom glass surfaces. These are for four different cases of MC concentrations (time interval: 100 sec, total observation time 10 min). **(b)** Mean orientation, obtained from Gaussian curves fitting to each histogram in (a) (excluding the case of no MC), plotted as a function of time. **(c)** Mean orientation, obtained from simulations using three different alignment interaction strengths, $\alpha$, for $N=256$, $M=32$, $\rho=0.44$, spontaneous curvature $\theta_0=10^{-3}$, and bending rigidity $\kappa=5$. Angle and time are translated from simulation values (see Materials and Methods). **(d)** Variance of the orientation distribution plotted as a function of time. See Materials and Methods for the detailed procedure.

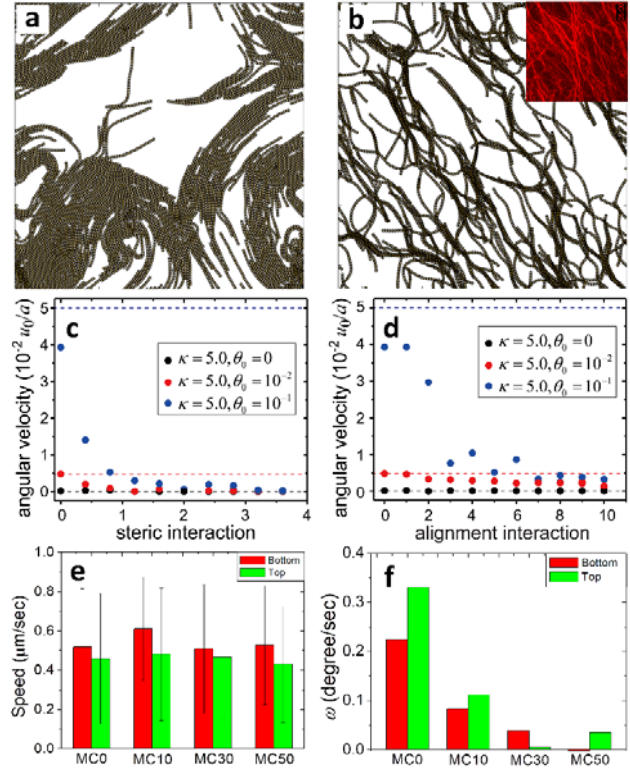

**Figure 4**. Speed and angular velocity of gliding MTs, modeling analysis and simulation: **(a, b)** Snapshots of the simulations for the steric interactions (a) and for the alignment interactions (b) $N = 256$, $M = 32$, $\rho = 0.44$ for bending rigidity $\kappa = 5$. Inset in panel (b) indicates a fluorescence image of aligned MTs (the second panel of middle row in Fig. 2(a), rotated counterclockwise by 90°) for comparison. **(c, d)** Mean angular velocity for $N = 256$, $M = 32$, $\rho = 0.44$ for bending rigidity $\kappa = 5$, and various spontaneous curvatures with steric interactions (c) and with alignment interactions (d). The dashed lines indicate the angular velocity for an isolated filament. **(e)** Mean gliding speed dependent on MC concentration. Red/Green bars are results from experimental observations of bottom/top surfaces, respectively. Three different areas were averaged for each surface. Vertical lines indicate standard deviations. **(f)** Experimentally observed mean angular velocity ($\omega$) dependent on MC concentration. The resulting values were multiplied by minus one for the case of the bottom surface, such that positive values in this plot indicate counterclockwise rotation. See Materials and Methods for detailed description.

## ASSOCIATED CONTENT

**Supporting Information**

The following files are available free of charge. Supplementary Figures (PDF), and Supplementary Movies (AVI): Active network aligned on a bottom surface of a flow cell (Supplementary Movie 1), Active network aligned on a top surface of a flow cell (Supplementary Movie 2), Active network with diluted visible MTs aligned on a bottom surface of a flow cell (Supplementary Movie 3), Active network with diluted visible MTs aligned on a top surface of a flow cell (Supplementary Movie 4), Simulation result of gliding MTs with an alignment interaction (Supplementary Movie 5), Simulation result of gliding MTs with a steric interaction (Supplementary Movie 6), Gliding MTs (Supplementary Movie 7).

## AUTHOR INFORMATION

**Corresponding Author**

*E-mail: teizer@tamu.edu

**Author Contributions**

K.K and N.Y. designed the experiments and analyzed the data. K.K., S.B. and H.N. performed the experiments. N.Y. performed the theoretical modeling. K.K., N.Y. and W.T. wrote the paper. M.U. and W.T. directed the research. All authors reviewed the paper.

**Funding Sources**

Japan Society for the Promotion of Science (JSPS) KAKENHI Grants JP26800219, JP17K05605, and JP16H00793.

ACKNOWLEDGMENT

We gratefully acknowledge support from the World Premier International Research Center Initiative (WPI), MEXT, Japan and the Fusion Research Program of WPI-AIMR. The authors are grateful to Sakurako Tanida and Tetsuya Hiraiwa for helpful discussions.

REFERENCES

(1) Alberts, B.; Johnson, A.; Lewis, J.; Raff, M.; Roberts, K.; Walter, P. Molecular Biology of The Cell, 4th Edition, Garland Science, New York

(2) Brown, N. A.; Wolpert, L. *Development* **1990**, 109, 1-9

(3) Henley, C. L. *J. Stat. Phys*. **2012**, 148, 741-775

(4) Pohl, C. *Symmetry* **2015**, 7, 2062-2107

(5) Danilchik, M. V.; Brown, E. E.; Riegert, K. *Development* **2006**, 133, 4517-4526

(6) Tee, Y. H.; Shemesh, T.; Thiagarajan, V.; Hariadi, R. F.; Anderson, K. L.; Page, C.; Volkmann, N.; Hanein, D.; Sivaramakrishnan, S.; Kozlov, M. M.; Bershadsky, A. D. *Nat. Cell Biol*. **2015**, 17, 445-457

(7) Naganathan, S. R.; Middelkoop, T. C.; Fürthauer, S.; Grill, S. W. *Curr. Opin. Cell Biol*. **2016**, 38, 24-30

(8) Ray, S.; Meyhöfer, E.; Milligan, R. A.; Howard, J. *J. Cell Biol*. **1993**, 121, 1083-1093

(9) Liu, H.; Spoerke, E. D.; Bachand, M.; Koch, S. J.; Bunker, B. C.; Bachand, G. D. *Adv. Mater*. **2008**, 20, 4476-4481

(10) Kawamura, R.; Kakugo, A.; Shikinaka, K.; Osada, Y.; Gong, J. P. *Biomacromolecules* **2008**, 9, 2277-2282

(11) Nitzsche, B.; Ruhnow, F.; Diez, S. *Nat. Nanotechnol.* **2008**, 3, 552-556

(12) Kakugo, A.; Kabir, A. Md. R.; Hosoda, N.; Shikinaka, K.; Gong, J. P. *Biomacromolecules* **2011**, 12, 3394-3399

(13) Furukawa, H.; Gong, J. P.; Masunaga, H.; Masui, T.; Koizumif, S.; Shikinaka, K. *Soft Matter* **2012**, 8, 11544-11551

(14) Shikinaka, K.; Mori, S.; Shigehara, K.; Masunaga, H. *Soft Matter* **2015**, 11, 3869-3874

(15) Lobikin, M.; Wang, G.; Xu, J.; Hsieh, Y.-W.; Chuang, C.-F.; Lemire, J. M.; Levin, M. *Proc. Natl. Acad. Sci. U.S.A.* **2012**, 109, 12586-12591

(16) Thitamadee, S.; Tuchihara, K.; Hashimoto, T. *Nature* **2002**, 417, 193-196

(17) McDowell, G. S.; Lemire, J. M.; Paré, J.-F.; Cammarata, G.; Lowery, L. A.; Levin, M. *Integr. Biol.* **2016**, 8, 267-286

(18) Inoue, D.; Mahmot, B.; Kabir, A. Md. R.; Farhana, T. I.; Tokuraku, K.; Sada, K.; Konagaya, A.; Kakugo, *Nanoscale* **2015**, 7, 18054-18061

(19) Nishiguchi, D.; Nagai, K. H.; Chaté, H.; Sano, M. *Phys. Rev. E* **2017**, 95, 020601(R)

(20) Rezakhaniha, R.; Agianniotis, A.; Schrauwen, J. T. C.; Griffa, A.; Sage, D.; Bouten, C. V. C.; van de Vosse, F. N.; Unser, M.; Stergiopulos, N. *Biomech. Model. Mechanobiol.* **2012**, 11, 461-473

(21) de Gennes, P. G.; Prost, J. **1993** The physics of liquid crystals, Oxford University Press, UK.

(22) Cates, M. E.; Tailleur, J. *Annu. Rev. Condens. Matter Phys*. **2015**, 6, 219-244

(23) Fily, Y.; Marchetti, M. C. *Phys. Rev. Lett*. **2012**, 108, 235702

(24) Denk, J.; Huber, J.; Reithmann, E.; Frey, E. *Phys. Rev. Lett*. **2016**, 116, 178301

(25) Sumino, Y.; Nagai, K. H.; Shitaka, Y.; Tanaka, D.; Yoshikawa, K.; Chaté, H.; Oiwa, K. Nature **2012**, 483, 448-452

(26) Hyman, A. A.; Chrétien, D.; Arnal, I.; Wade, R. H. *J. Cell. Biol*. **1995**, 128, 117-125

(27) Wells, D. B.; Aksimentiev, A. *Biophys. J.* **2010**, 99, 629-637

(28) Forgacs, G. *J. Cell. Sci*. **1995**, 108, 2131-2143

(29) Elinson, R. P.; Rowning, B. *Dev. Biol*. **1998**, 128, 185-197

(30) Cha, B.-J.; Gard, D. L. *Dev. Biol*. **1999**, 205, 275-286

(31) Gerhart, J.; Danilchik, M.; Doniach, T.; Roberts, S.; Rowning, B.; Stewart, R. *Development* **1989**, 107, 37-51

(32) Houliston, E. *Development* **1994**, 120, 1213-1220

(33) Larabell, C. A.; Rowning, B. A.; Wells, J.; Wu, M.; Gerhart, J. C. *Development* **1996**, 122, 1281-1289

(34) Tran, L. D.; Hino, H.; Quach, H.; Lim, S.; Shindo, A.; Mimori-Kiyosue, Y.; Mione, M.; Ueno, N.; Winkler, C.; Hibi, M.; Sampath, K. *Development* **2012**, 139, 3644-3652

(35) Yuan, M.; Shaw, P. J.; Warn, R. M.; Lloyd, C. W. Proc. Natl. Acad. Sci. U.S.A. **1994**, 91, 6050-6053

(36) Chan, J.; Crowell, E.; Eder, M.; Calder, G.; Bunnewell, S.; Findlay, K.; Vernhettes, S.; Höfte, H.; Lloyd, C. *J. Cell Sci*. **2010**, 123, 3490-3495

(37) Pope, K. L.; Harris, T. J. C. *Development* **2008**, 135, 2227-2238

(38) Schaller, V.; Weber, C.; Semmrich, C.; Frey, E.; Bausch, A. R. *Nature* **2010**, 467, 73-77

(39) Köhler, S.; Schaller, V.; Bausch, A. R.; *Nat. Mater*. **2011**, 10, 462-468

(40) Sumino, Y.; Nagai, K. H.; Shitaka, Y.; Tanaka, D.; Yoshikawa, K.; Chaté, H.; Oiwa, K. *Nature* **2012**, 483, 448-452

(41) Sanchez, T.; Chen, D. T. N.; DeCamp, S. J.; Heymann, M.; Dogic, Z. *Nature* **2012**, 491, 431-435

(42) Surrey, T.; Nédélec, F.; Leibler, S.; Karsenti, E. *Science* **2001**, 292, 1167-1171

(43) Vignjevic, D.; Yarar, D.; Welch, M. D.; Peloquin, J.; Svitkina, T.; Borisy, G. G. *J. Cell Biol*. **2003**, 160, 951-962

(44) Sanchez, T.; Welch, D.; Nicastro, D.; Dogic, Z. *Science* **2011**, 333, 456-459

(45) Reymann, A.-C.; Boujemaa-Paterski, R.; Martiel, J.-L.; Guérin, C.; Cao, W.; Chin, H. F.; De La Cruz, E. M.; Théry, M.; Blanchoin, L. *Science* **2012**, 336, 1310-1314

(46) Keber, F. C.; Loiseau, E.; Sanchez, T.; DeCamp, S. J.; Giomi, L.; Bowick, M. J.; Marchetti, M. C.; Dogic, Z.; Bausch, A. R. *Science* **2014**, 345, 1135-1139

(47) Maloney, A.; Herskowitz, L. J.; Koch, S. J. *PLoS One* **2011**, 6, e19522

(48) Kim, K.; Liao, A.; Sikora, A.; Oliveira, D.; Nakazawa, H.; Umetsu, M.; Kumagai, I.; Adschiri, T.; Hwang, W.; Teizer, W. *Microdevices* **2014**, 16, 501–508

(49) Coy, D. L.; Wagenbach, M.; Howard, J. *J. Biol. Chem*. **1999**, 274, 3667–3671

(50) Sikora, A.; Canova, F. F.; Kim, K.; Nakazawa, H.; Umetsu, M.; Kumagai, I.; Adschiri, T.; Hwang, W.; Teizer, W. *ACS Nano* **2015**, 9, 11003-11013

(51) Hegge, S.; Kudryashev, M.; Smith, A.; Frischknecht, F. *Biotechnol. J.* **2009**, 4, 903-913

(52) Hyman, A. A.; Chrétien, D.; Arnal, I.; Wade, R. H. *J. Cell. Biol*. **1995**, 128, 117-125

(53) Mickey, B.; Howard, J. *J. Cell. Biol*. **1995**, 130, 909-917

(54) Nam, G.-M.; Lee, N.-K.; Mohrbach, H.; Johner, A.; Kulić, I. M. *Eur. Phys. Lett*. **2012**, 100, 28001

(55) Gosselin, P.; Mohrbach, H.; Kulić, I. M.; Ziebert, F. *Physica D* **2016**, 318, 105-111

(56) Gittes, F.; Mickey, B.; Nettleton, J.; Howard, J. *J. Cell Biol*. **1993**, 120, 923-923

(57) Vicsek, T.; Czirók, A.; Ben-Jacob, E.; Cohen, I.; Shochet, O. *Phys. Rev. Lett*. **1995**, 75, 1226-1229

(58) Ginelli, F.; Peruani, F.; Bär, M.; Chaté, H. *Phys. Rev. Lett*. **2010**, 104, 184502

(59) Ngo, S.; Peshkov, A.; Aranson, I. S.; Bertin, E.; Ginelli, F; Chaté, H. *Phys. Rev. Lett*. **2014**, 113, 038302

(60) Peruani, F. *Eur. Phys. J. ST* **2016**, 225, 2301-2317

(61)  Doi, M.; Edwards, S. The Theory of Polymer Dynamics, Clarendon, **1986**

**Supplementary Figure 1**

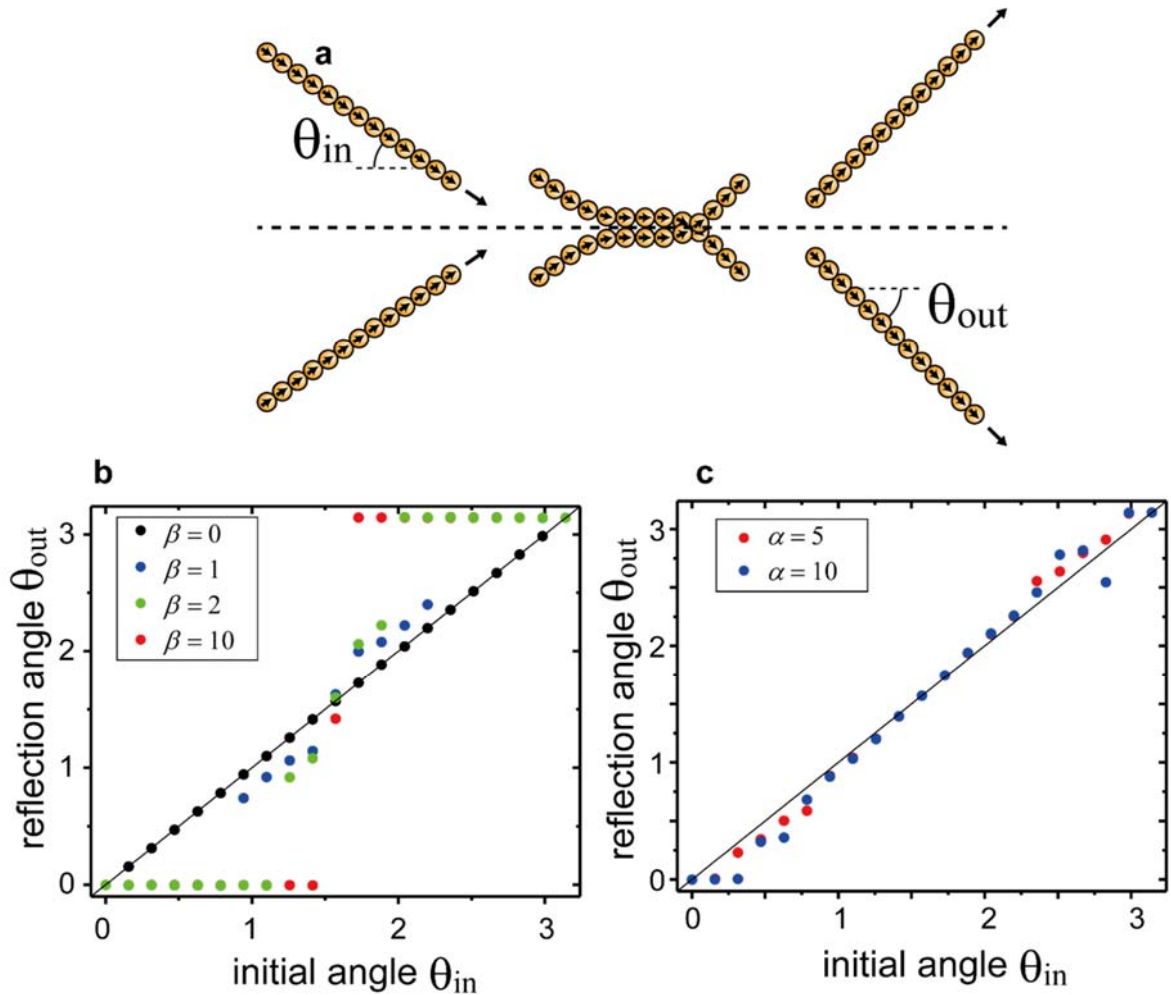

**Supplementary Fig. 1:** Collisions of two filaments in our model. (a) Schematic images of filament conformation during the collision. (b,c) Reflection angles as a function of initial angles of the collision to characterise the degree of alignment ability. The steric interactions (b) and the alignment interactions (c) were used. The solid lines indicate that the reflection angles are unchanged from the initial angles during the collisions. Deviation from the solid line implies an inelastic collision with alignment if the reflection angles are below the line for initial angles less than $\pi/2$, and if they are above the line for initial angles larger than $\pi/2$. The opposite case indicates inhibition of alignment. Both interactions show alignment. Deviation for the steric interactions is larger, and this demonstrates the steric interactions have stronger alignment ability.

**Supplementary Figure 2 (caption next page)**

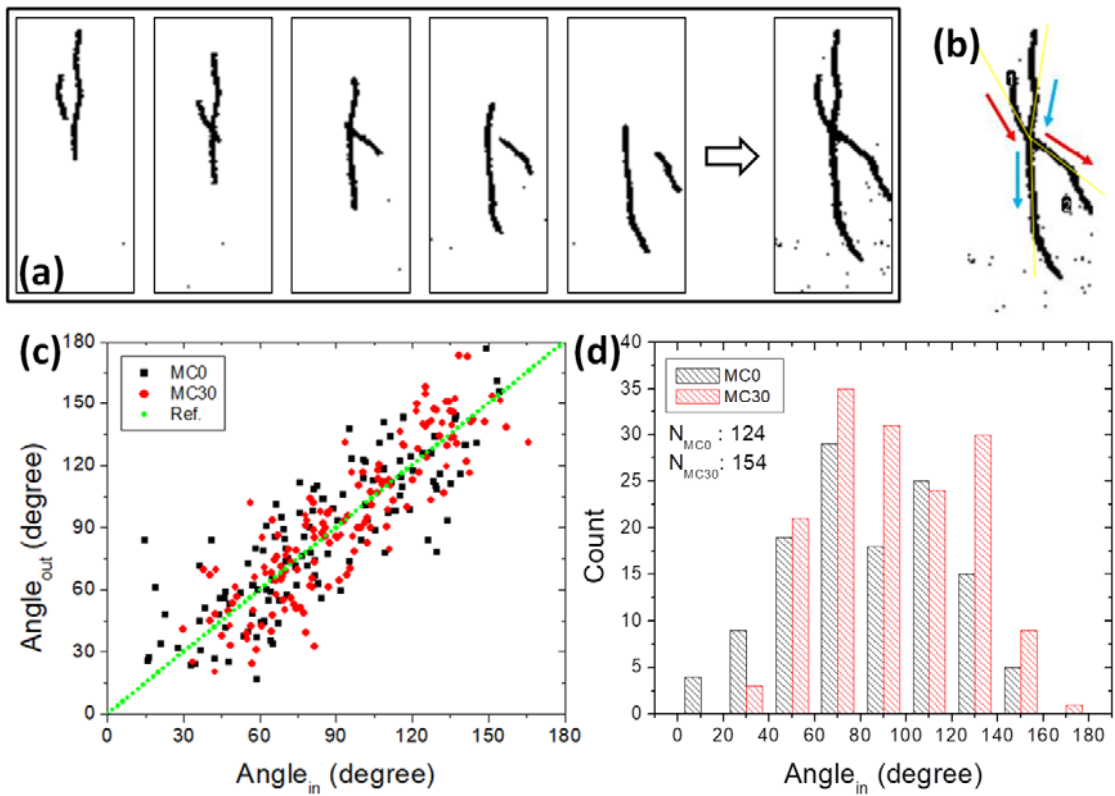

**Supplementary Fig. 2**: Analysis of MT crossing events. (a) Fluorescence images (binary format) of two gliding MTs crossing each other with time. The five panels are stacked up in the far right panel (right of the arrow) generating a trace of the gliding MTs. (b) Incident angle ((1), $Angle_{in}$) and outgoing angle ((2), $Angle_{out}$) denoted in the trace with yellow lines. Color-coded arrows follow each of the two MTs. (c) Scatter plot of $Angle_{in}$ versus $Angle_{out}$ for two different cases, with (MC30) and without (MC0) methyl cellulose. The reference line indicates $Angle_{in}/Angle_{out}=1$. (d) Incident angle histogram of MT crossing events. Total events: 124 for MC0 (three independent assays, four different areas for each assay) and 154 for MC30 (three independent assays, four different areas for each assay).

**Supplementary Figure 3**

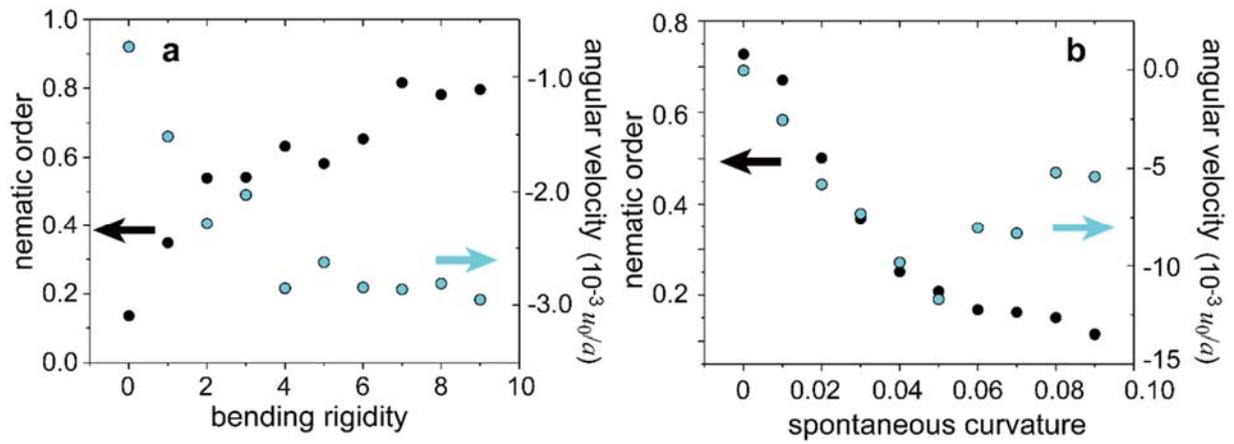

**Supplementary Fig. 3:** The nematic order parameter (black circles) and the angular velocity (light blue circles) dependence on bending rigidity (a) and spontaneous curvature (b).

**Supplementary Figure 4**

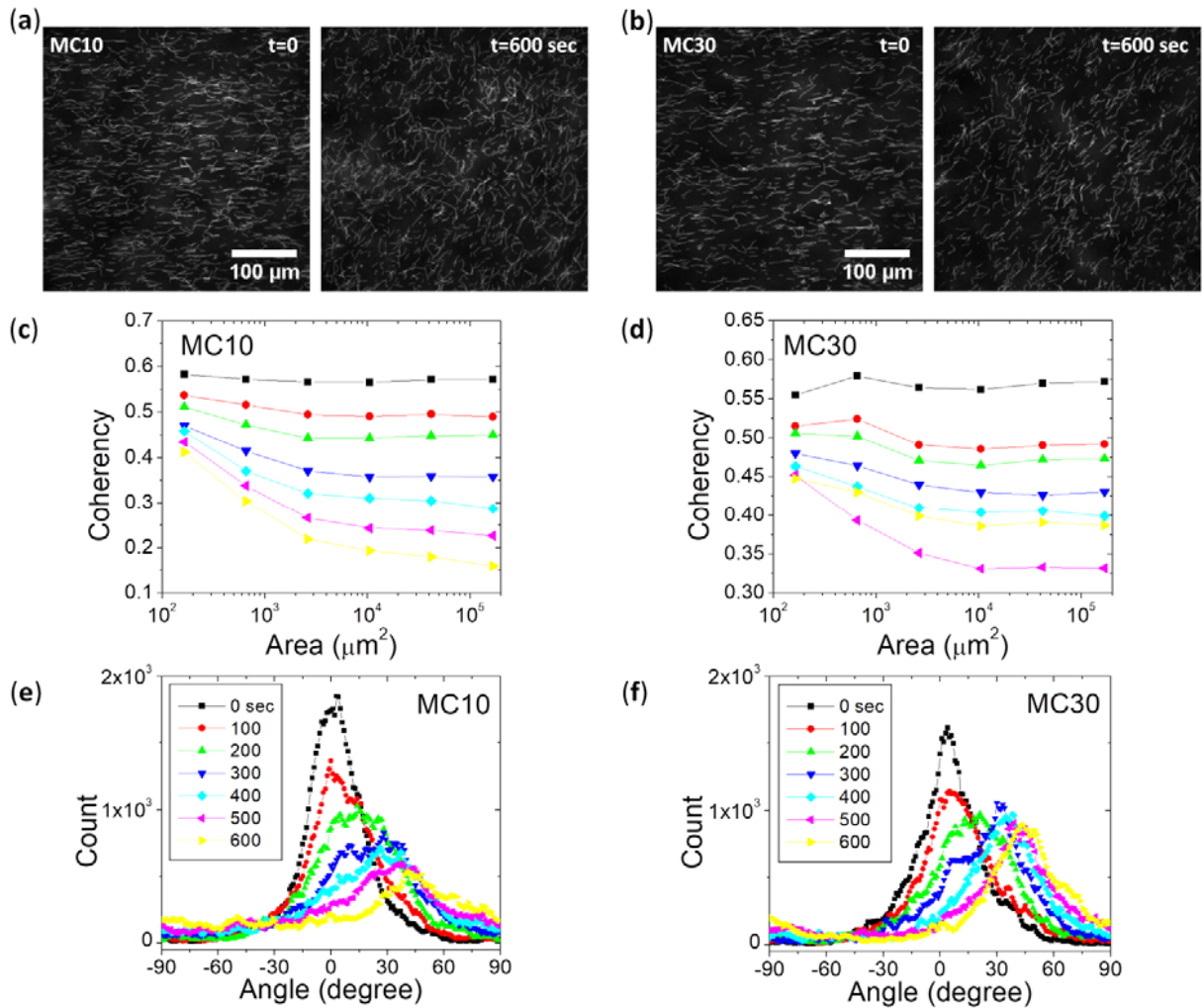

**Supplementary Fig. 4:** Coherency of MT orientation and local orientation histogram in the active layer formed in the presence of MC at two different concentrations, MC10 **(a, c, e)** and MC30 **(b, d, f)**. (a, b): time-lapsed fluorescence images (two different time frames, 0 and 600 sec) showing diluted visible MTs in the active layers. Scale bars: 100 μm. (c, d): Coherency plotted as a function of ROI size on a log-scale. (e, f): histograms of the local orientations of MTs plotted with absolute count. The zero-degree angle means the orientation parallel to the horizontal axis on each image and the orientation rotates counterclockwise as the angle increases. **Discussion**: In general, the coherency increases with the MC concentration and decreases with time. At lower MC concentration, the coherency drops faster (compare the curves for MC10 with those of MC30). The two cases show the coherency generally that decreases rapidly with the ROI size at the small area regime and saturates as the area increases further. The saturation level is lower for the case of MC10.

**Supplementary Figure 5**

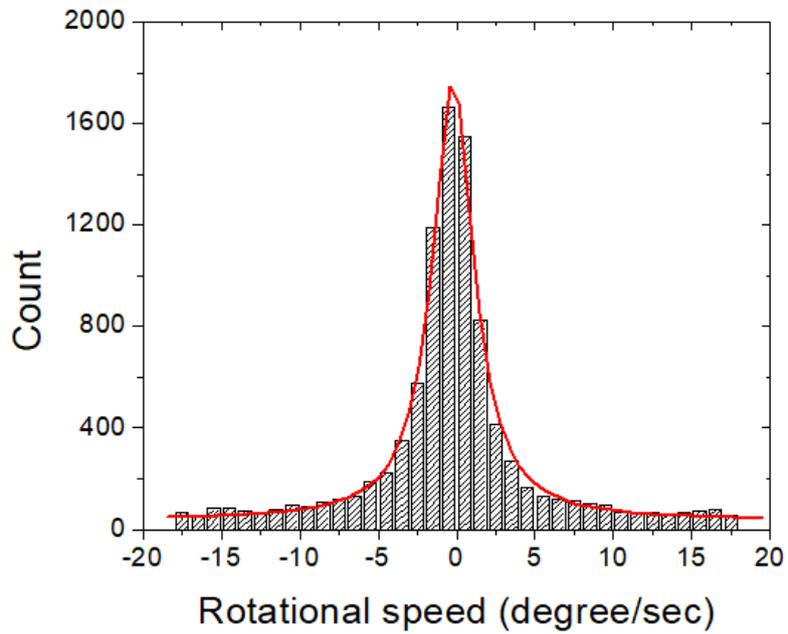

**Supplementary Fig. 5:** An example of the angular velocity (rotational speed) histogram and the Lorentzian curve fitting (red curve). The rotational speed is in degree/sec and this particular example shows the peak position ($x_c$) of −0.225 degree/sec.

**Supplementary Figure 6**

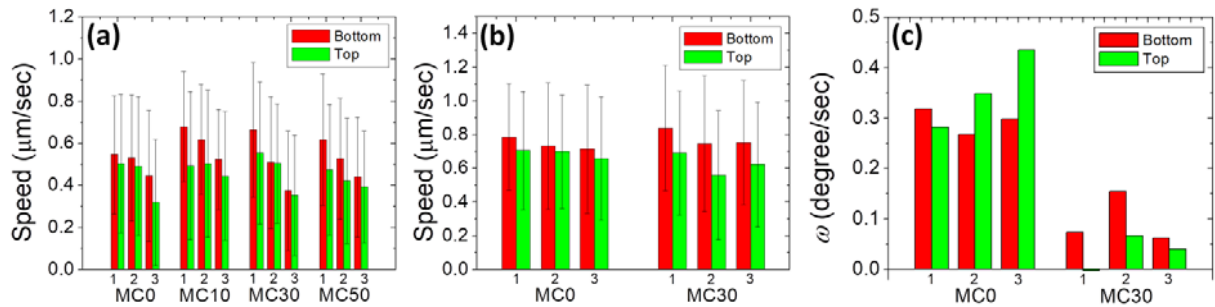

**Supplementary Fig. 6: (a)** MT gliding speed for four different MC concentrations. Red/green bars indicate results from bottom/top surface observation, respectively. The numbers, 1~3, for each set of the plot indicate the time sequence. Vertical lines are standard deviations. **(b-c)** MT gliding speed ((b)) and angular velocity $\omega$ ((c)) for two different MC concentrations. Red/green bars indicate results from bottom/top surface observation, respectively. The numbers, 1~3, for each set of the plot indicate three independent assays. Vertical lines in (b) are standard deviations. Here red bars in (c) are after multiplication of minus one, such that positive $\omega$ in this plot indicates counterclockwise rotation irrespective of the surface.

**Supplementary Figure 7**

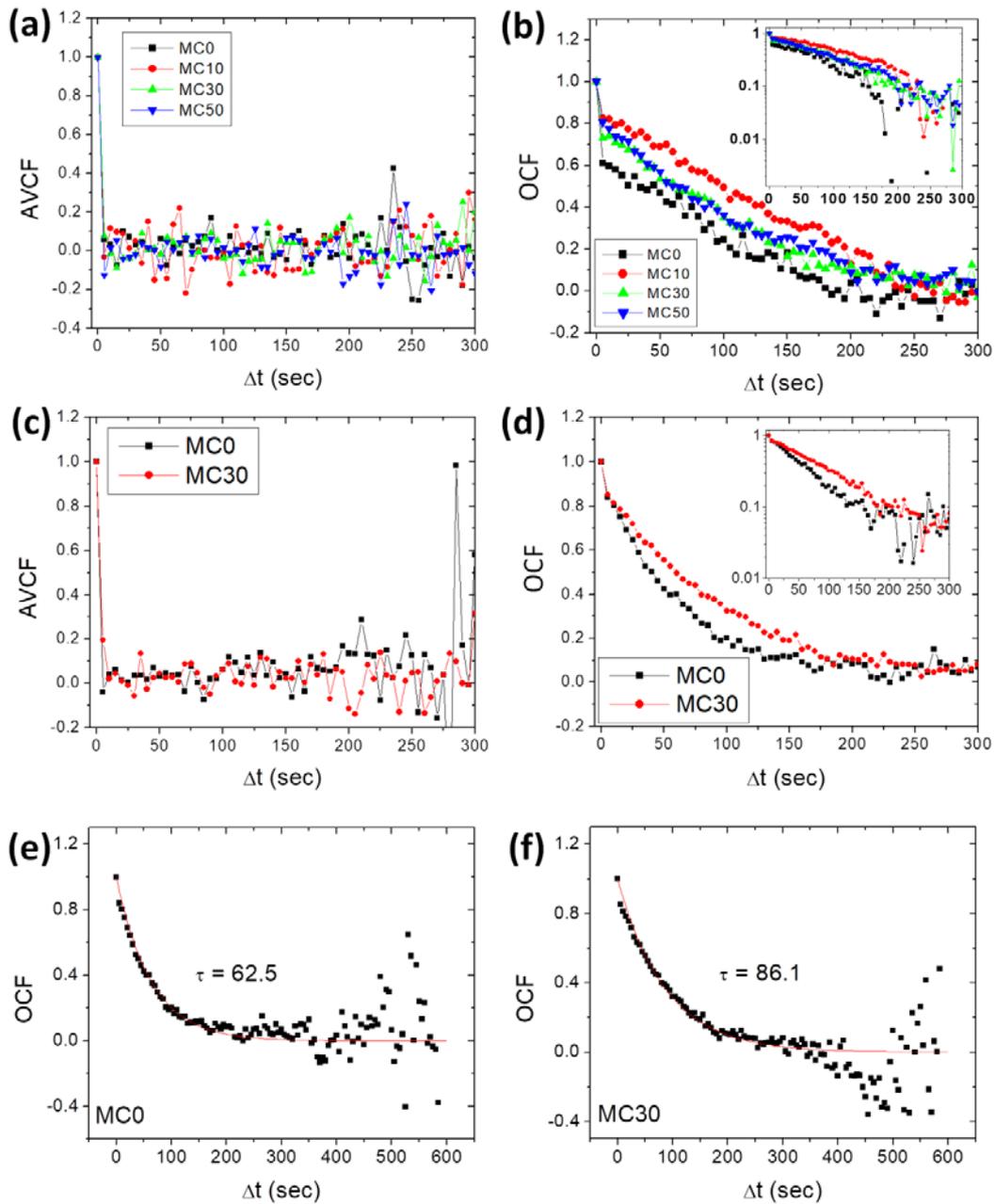

**Supplementary Fig. 7: (a)** Angular velocity correlation function (AVCF) and **(b)** orientation correlation function (OCF) as a function of lag time (Δt), obtained from the TOAST data tracing isolated gliding MTs. These plots show the assays analyzed for Fig. 4 (e, f) and Supplementary Fig. 6(a) for four different MC concentrations. The inset in panel (b) shows the OCFs plotted on a log-scale. AVCF drops immediately with Δt while OCF shows gradual decays with Δt. Data fluctuate more at higher Δt because the number of MTs traced decreases with increasing Δt. **(c)** Angular velocity correlation function (AVCF) and **(d)** orientation correlation function (OCF) as a function of lag time (Δt), obtained from the TOAST data tracing isolated gliding MTs. These plots show the assays analyzed for Supplementary Fig. 6(b-c) for two different MC concentrations. The inset in panel (d) shows the OCFs plotted on a log-scale. **(e, f)** Exponentially decaying curves fitting (red lines) to the OCFs, using a single parameter exponential function, $y = e^{-t/\tau}$. $\tau$ in each panel indicates the relaxation time in sec.

**Supplementary Figure 8** (caption next page)

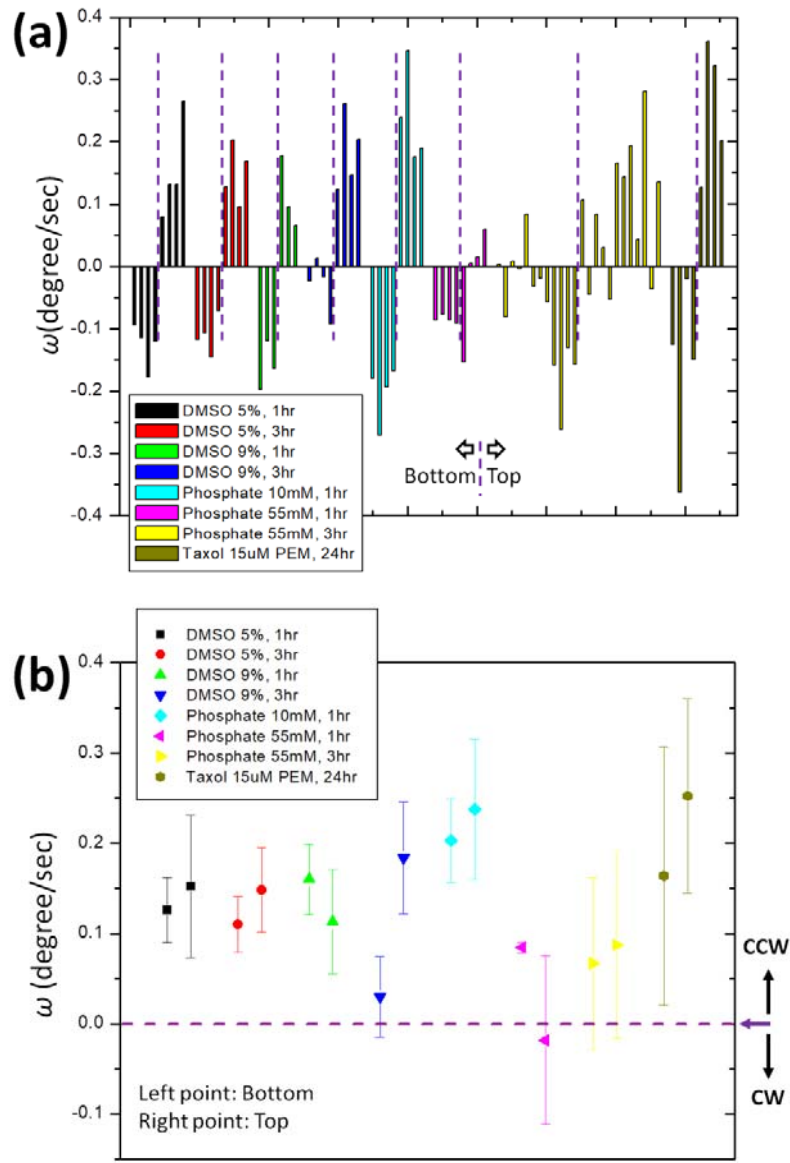

**Supplementary Fig. 8: (a)** Bar graph indicating mean angular velocities of MTs polymerized in various different conditions. At least two independent assays were performed for each case. Each bar is a result from an independent observation. Bars on the left side of vertical dotted lines (purple colored) are results from the bottom surface of the flow cell chamber, while bars on the right side are results from the top surface. These data show the measured values as it is. Therefore a negative value in the case of the bottom surface means counterclockwise rotation, while it indicates clockwise rotation for the case of the top surface. Preference of the counterclockwise rotation is apparent on both surfaces. For the cases of DMSO, the legend indicates DMSO concentration in % (in PEM buffer) and incubation time. Tubulin concentration

during polymerization was 2.5 mg/ml. For the cases of Phosphate, the legend indicates phosphate concentration and incubation time. Tubulin concentration during polymerization was 2.5 mg/ml. Note that 2 μM taxol was included in the phosphate buffer solutions. For the case of Taxol, PEM buffer including 15 μM taxol was used, the incubation time was 24 hr, and the tubulin concentration during polymerization was 0.5 mg/ml. All the buffer solutions included 1 mM GTP and 4 ~ 6 mM $MgCl_2$. **(b)** Averages with standard deviations (vertical lines) of the data in (a) taken for each case (color coded). The left/right data point in each set means the average for the case of bottom/top surface. Data from the bottom surface was multiplied by -1. As a result, a positive value in this graph indicates counterclockwise rotation irrespective of the surface.

**Supplementary Figure 9**

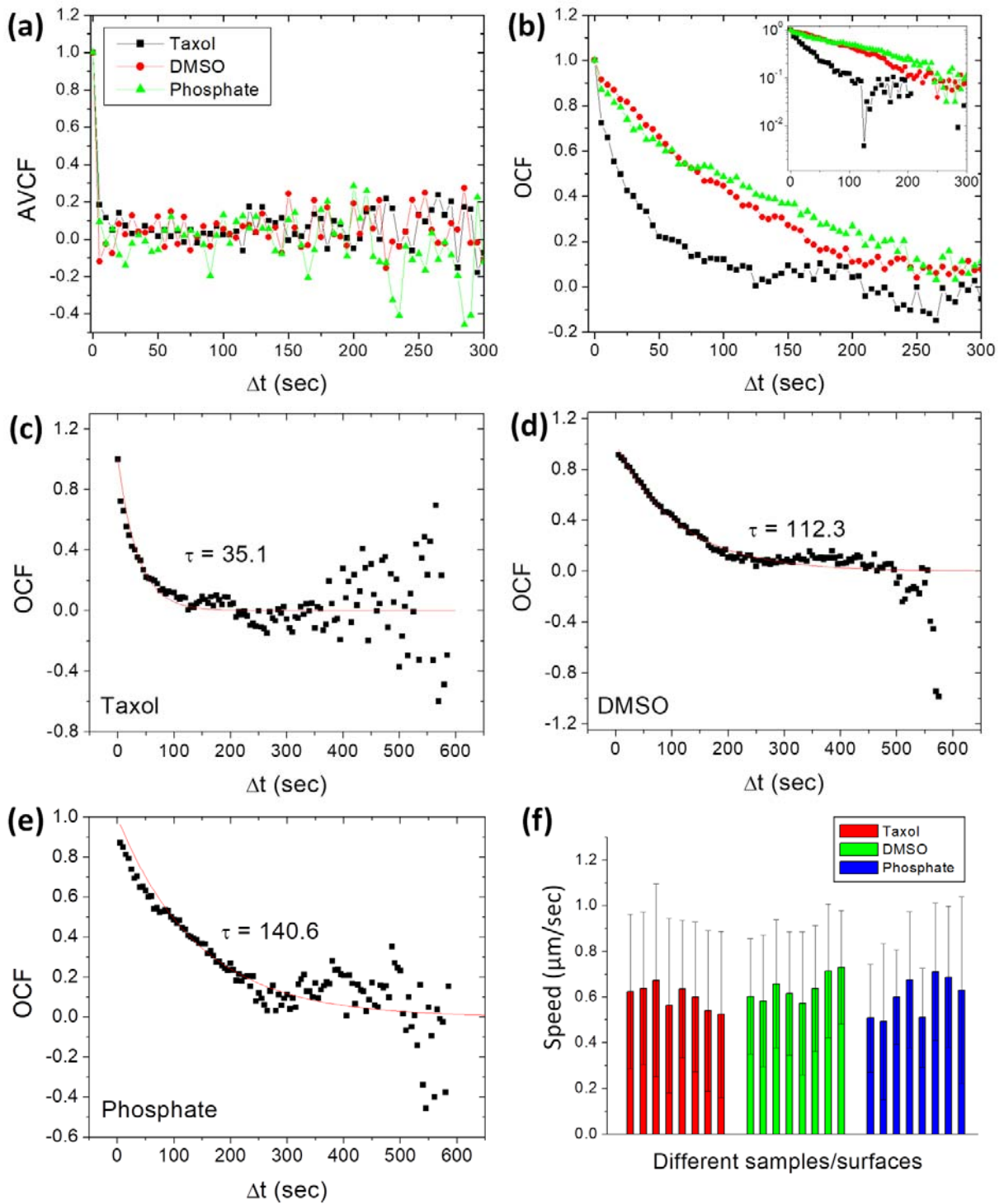

**Supplementary Fig. 9:** Three cases among all the conditions in Supplementary Fig. 8 were taken for the correlation analysis. These three conditions are (1) PEM buffer with 15 μM taxol

(Taxol), (2) PEM with 5 % DMSO, 1hr polymerization (DMSO) and (3) 55 mM Phosphate buffer with 2 µM taxol, 1 hr polymerization (Phosphate). **(a)** Angular velocity correlation function (AVCF) as a function of lag time (Δt). **(b)** Orientation correlation function (OCF) as a function of Δt. (Inset: the same plot on a log-scale). **(c-e)** Exponentially decaying curves fitting (red lines) to the OCFs in (b), using a single parameter exponential function, $y = e^{-t/\tau}$. $\tau$ in each panel indicates the relaxation time in sec. **(f)** Mean MT gliding speeds for the three selected tubulin polymerization conditions. Each bar is a result from an independent observation (see Supplementary Fig. 8a).

**Supplementary Figure 10**

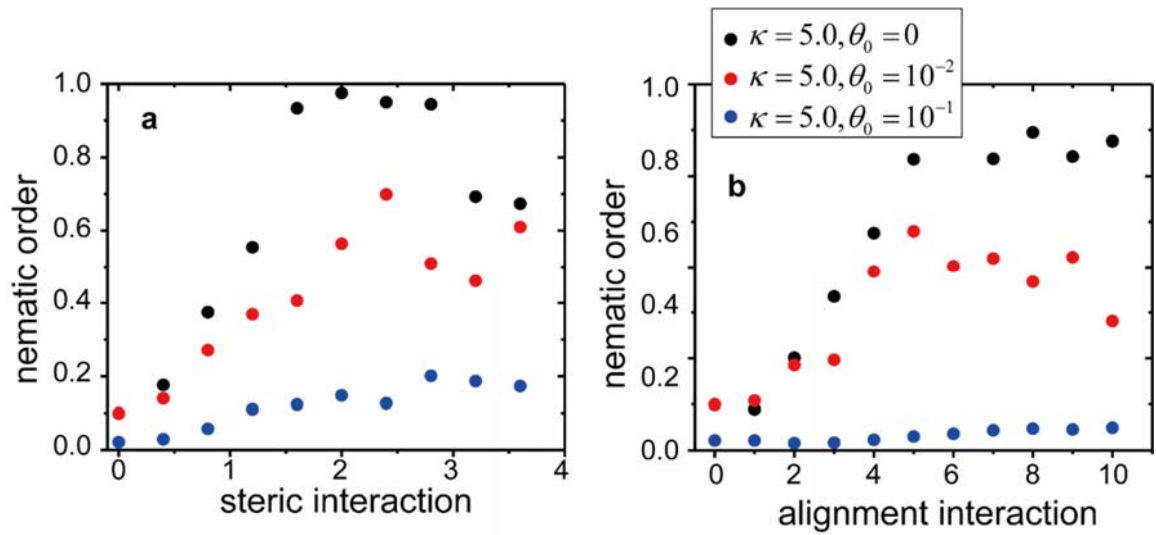

**Supplementary Fig. 10:** The nematic order parameter for $N = 256$ filaments of $M = 32$ monomers each, density $\rho = 0.44$, bending rigidity $\kappa = 5$ and various spontaneous curvatures $\theta_0$ with steric interactions, $\beta$ (a), and with alignment interactions, $\alpha$ (b), respectively.